\documentclass{article}
\usepackage{spconf,amsmath,graphicx,hyperref}
\usepackage{amsmath}   
\usepackage{amssymb}   
\usepackage{amsfonts}  
\usepackage{multirow}
\usepackage{bm} 
\usepackage{graphicx}  
\usepackage{booktabs}  
\usepackage{cite}      

\title{Fusion of Spatio-Temporal and Multi-Scale Frequency Features\\
for Dry Electrodes MI-EEG Decoding}

\name{Tianyi Gong$^{1 \dag}$, 
      Can Han$^{2 \dag}$,
      Junxi Wu$^{3}$,
      Dahong Qian$^{2 }$
\thanks{$\dag$These authors contributed to the work equally.},
\thanks{Partially supported by Shanghai AIMIC Medical Electronics Co.}%
      }

\address{$^{1}$ School of Data Science, The Chinese University of Hong Kong, Shenzhen, China \\
        $^{2}$ School of Biomedical Engineering, Shanghai Jiao Tong University, Shanghai, China \\
        $^{3}$ OYMotion Technology Co., Ltd \\
         122090863@link.cuhk.edu.cn, \{hancan, dahong.qian\}@sjtu.edu.cn, JCjunxi@gmail.com}

\begin{document}
%
\maketitle

\begin{abstract}
Dry-electrode Motor Imagery Electroencephalography (MI-EEG) enables fast, comfortable, real-world Brain Computer Interface by eliminating gels and shortening setup for at-home and wearable use.
However, dry recordings pose three main issues: lower Signal-to-Noise Ratio with more baseline drift and sudden transients; weaker and noisier data with poor phase alignment across trials; and bigger variances between sessions. These drawbacks lead to larger data distribution shift, making features less stable for MI-EEG tasks.
To address these problems, we introduce STGMFM, a tri-branch framework tailored for dry-electrode MI-EEG, which models complementary spatio-temporal dependencies via dual graph orders, and captures robust envelope dynamics with a multi-scale frequency mixing branch, motivated by the observation that amplitude envelopes are less sensitive to contact variability than instantaneous waveforms. Physiologically meaningful connectivity priors guide learning, and decision-level fusion consolidates a noise-tolerant consensus. On our collected dry-electrode MI-EEG, STGMFM consistently surpasses competitive CNN/Transformer/graph baselines.
Codes are available at \url{https://github.com/Tianyi-325/STGMFM}.
\end{abstract}

%

\section{Introduction}
Non-invasive Electroencephalography (EEG) is the most accessible neural sensing modality for daily-life Brain Computer Interface (BCI) applications. Among non-invasive options, dry-electrode systems avoid conductive gel, shorten setup/cleanup, and better fit out-of-lab use~\cite{lopezgordo2014dry,grummett2015dry}. This practicality directly makes it suitable for rehabilitation at home, longitudinal self-training and wearable assistive interaction, where user comfort, compliance, and scalability are critical. However, dry recordings differ systematically from wet data: the unstable skin–electrode coupling often elevates and fluctuates impedance, lowering Signal-to-Noise (SNR) and amplifying motion/contact artifacts~\cite{lopezgordo2014dry,grummett2015dry}. Besides, re-positioning across days leads to cross-session drift. What is more, inter-subject variability is exacerbated due to hair and pressure differences. In practice, dry-cap protocols also limit the number and duration of trials to reduce discomfort and time burden, which increases the premium on data-efficient and noise-robust decoding and makes frequent recalibration impractical. Currently, most existing Motor Imagery (MI) decoders were designed/validated under wet-electrode assumptions, and most of them transfer poorly to dry conditions due to the noise characteristics, distribution shift, and trial scarcity~\cite{lotte2018review}. As a result, explicit algorithmic tailoring for dry-electrode MI remains under-explored.

Non-invasive BCIs with dry electrodes ease setup, but they typically suffer lower SNR and contact variability, which depresses MI accuracy and calls for noise-robust, data-efficient models \cite{lopezgordo2014dry,grummett2015dry,kim2025cnn_dry_eeg}. Classical MI pipelines relied on Common Spatial Pattern/bandpower and linear classifiers or Riemannian geometry on covariances \cite{ramoser2000csp,ang2008fbcsp,barachant2013riemann,lotte2018review}, while deep CNNs improved end-to-end decoding yet still assume relatively clean, well-calibrated signals \cite{lawhern2018eegnet,schirrmeister2017deep,Han2024SSTDPN,ingolfsson2020eegtcnet,song2022eegconformer,miao2023lmdanet}. Graph formulations encode inter-channel relations and have shown benefits for spatio–temporal fusion and cross-session robustness \cite{zhang2022stgcn_eeg}—most notably ST-GF by Wang \emph{et al.} \cite{wang2024stgf}; multi-branch/domain-generalization schemes further diversify representations via decision fusion \cite{du2023multiview}. In time-series modeling, recent “mixers’’ learn multi-period structure with compact, multi-resolution tokenization, including TimeMixer and its successor TimeMixer++ by Wang \emph{et al.} \cite{wang2024timemixer,Wang2025TimeMixerPP}, as well as related sequence learners \cite{nie2023patchtst,wu2023timesnet}. Connectivity priors such as PLV provide phase-synchrony graphs that are amplitude-invariant and physiologically meaningful \cite{lachaux1999measuring,Kang2026PLVGCN}.

\begin{figure*}[t]
  \centering
  \includegraphics[width=\linewidth]{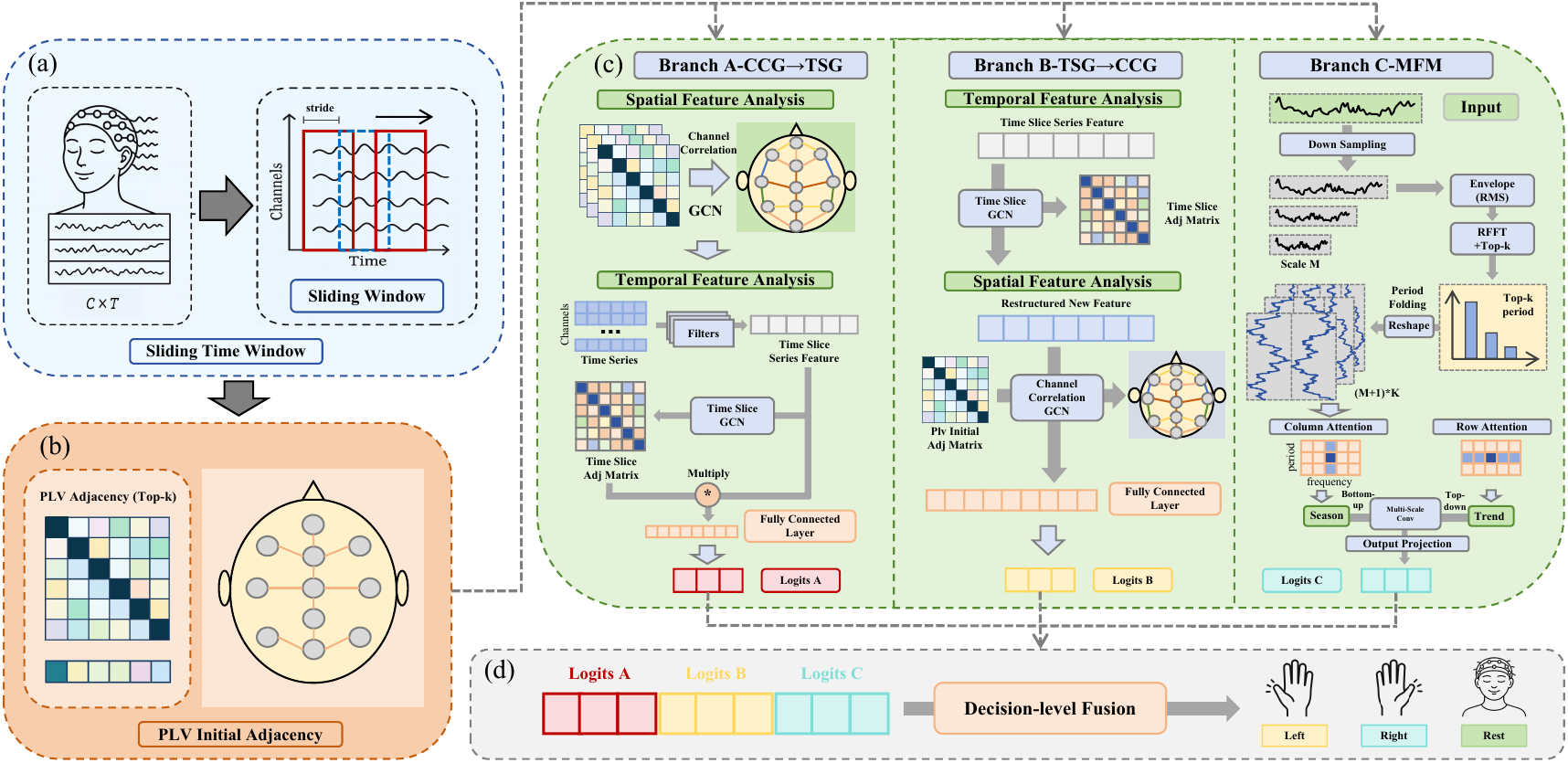}
  \caption{Pipeline of the proposed model. Raw EEG is windowed into overlapping segments and processed by three branches. \textbf{A} (CCG$\rightarrow$TSG) aligns functional connectivity via a PLV-initialized channel graph before aggregating on a time-slice graph. \textbf{B} (TSG$\rightarrow$CCG) first reconstructs intra-trial continuity and then fuses channels on the CCG. \textbf{C} (MFM) operates on the amplitude envelope (RMS) with multi-resolution imaging and cross-scale mixing. Linear head fuses branch logits at the decision level.}
  \label{fig:pipeline}
\end{figure*}

Although these advances are compelling, most were devised and tuned with wet-electrode data rather than dry EEG. Therefore, they fail to remediate the low-SNR, contact-motion-artifact instability. They also emphasize localized patterns and lack multi-view learning, limiting the learning of deep, invariant features to bridge cross-session/subject shifts. In addition, multi-scale spatio and temporal structure is under-explored, leaving deep cross-scale cues untapped.
Motivated by these gaps, we propose \textbf{STGMFM}, which unifies dual ordered spatio-temporal graph branches (CCG$\!\rightarrow\!$TSG and TSG$\!\rightarrow\!$CCG) so that spatial-first and temporal-first propagation can hedge against different noise pathways. And we augments them with a multi-cale frequency mixer that learns multi-resolution envelope dynamics, aligned with ERD/ERS to offer a robust temporal cue when instantaneous waveform detail is unreliable. Physiological connectivity priors (PLV) provide a relatively detailed and precise starting point that is then adapted end-to-end to accommodate subject/session variability, while decision-level fusion forms a simple consensus that resists over-reliance on any single branch. In experiments on a dry-electrode MI-EEG dataset, STGMFM outperforms CNN/transformer/graph baselines, proving it to be a practical path toward dry-electrode BCIs.

\textbf{Main contributions}:
(1) We introduce a PLV-initialized EEG graph that creates a more detailed and precise prior adjacency, suitable for low-SNR dry recordings.
(2) We design a dual-order spatio-temporal graph to yield complementary evidence.
(3) We introduce a lightweight Multi-Scale Frequency Mixer that extracts phase-invariant, multi-scale temporal cues aligned with ERD/ERS.
(4) We employ a stable training recipe (decision-level fusion, L1/L2 regularization, cosine annealing) that reduces overfitting and over-reliance.
\section{Methods}
\label{sec:methods}
\subsection{Overview}
We present a triple-branch framework for cross-subject MI EEG decoding. Raw trials are segmented with overlapping windows. Branches A/B pair a channel-correlation graph (CCG) and a time-slice graph (TSG) in opposite orders (CCG$\to$TSG vs.\ TSG$\to$CCG), yielding complementary biases—spatial alignment before temporal aggregation vs.\ temporal stabilization before spatial fusion. Branch C consumes amplitude envelopes to build a lightweight multi-resolution representation of rhythmic/trend dynamics. Each branch outputs logits, and a shallow head fuses them at the decision level, delivering robust generalization without fragile feature-level alignment. Figure~\ref{fig:pipeline} illustrates the overall pipeline.

\subsection{Notation and Sliding Windowing}
\label{subsec:notation}
Let a trial be $\mathbf{X}\in\mathbf{R}^{C\times T}$ with $C$ channels and $T$ time steps. Using window length $W_n$ and stride $\mathrm{Str}$, the number of windows is
\begin{equation}
  W=\left\lfloor\frac{T - W_n}{\mathrm{Str}}\right\rfloor + 1.
  \label{eq:W}
\end{equation}
After segmentation, we obtain a $4$-D tensor $\mathbf{X}\in\mathbf{R}^{N\times W\times C\times T_w}$ for batch size $N$ and per-window length $T_w=W_n$. Windowing provides natural nodes for the TSG and a consistent statistical unit for the graph modules and the envelope branch, avoiding drift on long unsegmented sequences.

\subsection{PLV-driven Initial Channel Graph}
\label{subsec:plv}
To encode functional connectivity as a prior, we build an initial adjacency from the phase-locking value (PLV). For channels $i$ and $j$ with analytic phases $\phi_i(t)$ and $\phi_j(t)$,
\begin{equation}
  \mathrm{PLV}(i,j)=\left|\frac{1}{T}\sum_{t=1}^{T} e^{\mathrm{i}\left(\phi_i(t)-\phi_j(t)\right)}\right|.
  \label{eq:PLV}
\end{equation}
We remove self-loops, sparsify by per-row Top-$k$ (or a threshold), symmetrize, and degree-normalize to obtain $\tilde{\mathbf{A}}=\mathbf{D}^{-1/2}\mathbf{A}\mathbf{D}^{-1/2}$. The same $\tilde{\mathbf{A}}$ initializes the CCG in both branches. During training, a small learnable increment adapts $\tilde{\mathbf{A}}$ to subject-specific variability while retaining interpretability. At the same time, it also provides more detailed and precise adjacency.
Fig.~\ref{fig:ccg_before_after} shows the before/after learning comparison of adjacency under PLV-initial and the basic spatial-neighborhood prior.

\subsection{Branch A: \texorpdfstring{CCG$\rightarrow$TSG}{CCG->TSG}}
\label{subsec:branchA}
\begin{figure}[t]
  \centering
  \includegraphics[width=\linewidth]{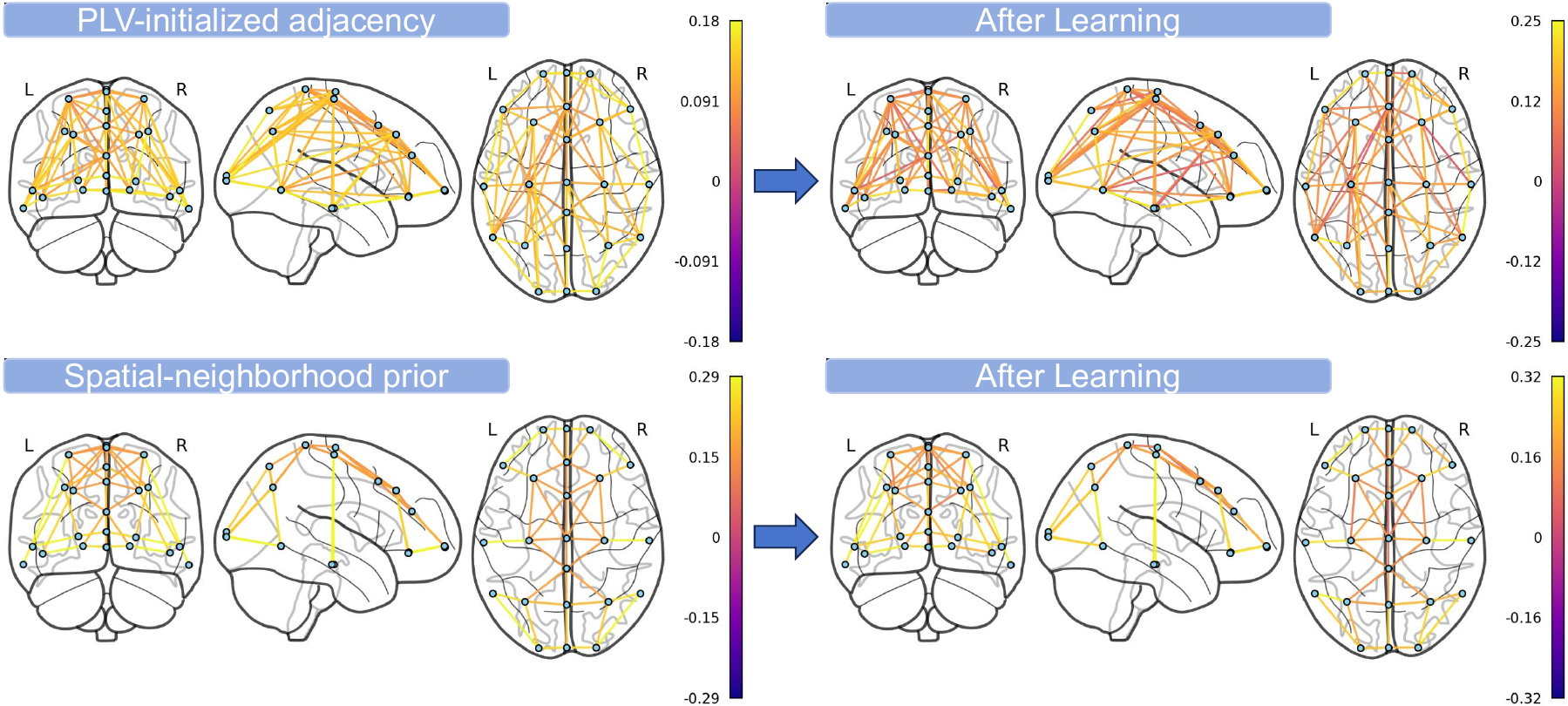}
  \caption{Before/after training comparison of adjacency under two priors: PLV initialization and the spatial-neighborhood prior}
  \label{fig:ccg_before_after}
\end{figure}

Branch A first performs graph propagation on the channel graph to align truly cooperating electrodes and suppress mismatched ones. With node features $\mathbf{H}^{(l)}\in\mathbf{R}^{C\times d}$, one CCG layer reads
\begin{equation}
  \mathbf{H}^{(l+1)}=\sigma\!\left(\tilde{\mathbf{A}}\;\mathbf{H}^{(l)}\;\mathbf{W}_s^{(l)}\right),\quad l=0,\dots,K_s-1.
  \label{eq:CCG}
\end{equation}
We visualized the changes in spatial connectivity before and after learning, see Fig.~\ref{fig:ccg_before_after}. The resulting spatially denoised representation then passes through a shared temporal convolutional block (depthwise–pointwise $1$D convolution with GELU/normalization) and is aggregated over the \emph{time-slice graph}. Let $\mathbf{U}^{(l)}\in\mathbf{R}^{W\times d}$ be slice-level features and $\tilde{\mathbf{S}}\in\mathbf{R}^{W\times W}$ the learnable slice adjacency; one TSG layer is
\begin{equation}
  \mathbf{U}^{(l+1)}=\sigma\!\left(\tilde{\mathbf{S}}\;\mathbf{U}^{(l)}\;\mathbf{W}_t^{(l)}\right),\quad l=0,\dots,K_t-1.
  \label{eq:TSG}
\end{equation}
Placing temporal aggregation on top of a connectivity-aligned representation improves the SNR of time reconstruction by reducing sensitivity to phase misalignment across channels. A global average pooling (GAP) and a linear head yield $\mathbf{z}_A$. We insert a lightweight shared temporal block between CCG and TSG to align feature spaces and add local temporal expressiveness without disturbing the graph inductive bias.

\subsection{Branch B: \texorpdfstring{TSG$\rightarrow$CCG}{TSG->CCG}}
\label{subsec:branchB}
Branch B starts by reconstructing intra-trial continuity on the slice graph. Continuous MI segments become tightly clustered while non-task fragments are attenuated. The graph-filtered features are then rearranged to the channel dimension and passed through the CCG propagation with $\tilde{\mathbf{A}}$. By stabilizing time first, B prevents short-term temporal noise from being prematurely diffused spatially. A shared temporal block and classifier head produce $\mathbf{z}_B$. Shared temporal block is also placed between TSG and CCG.

\subsection{Branch C: Multi-Scale Frequency Mixer}
\label{subsec:etm}
The third branch operates on the amplitude envelope (RMS) (e.g., $\mu/\beta$ band), which makes ERD/ERS modulation explicit and phase-invariant. We first form multi-resolution ``time images'' by selecting dominant periods/frequencies (rfft) and mapping the $1$D signal to a compact time–frequency lattice. We then decouple periodic (seasonal) patterns from slower trend evolution using dual-axis processing, and finally carry out multi-scale and multi-resolution mixing to aggregate evidence. This imaging—decoupling—mixing pipeline turns long dependencies into local ones on the image plane and yields a compact cross-scale representation. A lightweight configuration (small channel width and a single block, with one down-sampling and Top-1 dominant period) suits the low-sample EEG regime. The branch outputs $\mathbf{z}_C$.

\begin{table*}[t]
\centering
\caption{Avg$\pm$Std results on dry-EEG under three evaluation protocols. We report Accuracy (ACC), Cohen's kappa, and F1.}
\label{tab:merged_main}
\resizebox{\linewidth}{!}{%
\begin{tabular}{lccccccccc}\toprule
\multirow{2}{*}{Method} & \multicolumn{3}{c}{Cross Session} & \multicolumn{3}{c}{Cross Subject} & \multicolumn{3}{c}{Cross-Subject + Single-Session Fine-tuning} \\
\cmidrule(lr){2-4}\cmidrule(lr){5-7}\cmidrule(lr){8-10}
& ACC(\%) & kappa & F1(\%) & ACC(\%) & kappa & F1(\%) & ACC(\%) & kappa & F1(\%) \\
\midrule
ShallowNet~\cite{schirrmeister2017deep}        & $47.30\pm7.68$ & $0.2095\pm0.1154$ & $45.39\pm7.85$  & $50.38\pm10.92$& $0.2621\pm0.1643$  & $48.55\pm12.13$ & $52.06\pm11.30$& $0.2809\pm0.1695$ & $50.18\pm12.03$ \\
EEGNet~\cite{lawhern2018eegnet}                & $47.67\pm11.78$& $0.2152\pm0.1767$ & $45.96\pm12.41$ & $52.21\pm10.81$& $0.2936\pm0.1625$  & $50.17\pm11.85$ & $54.94\pm10.23$& $0.3241\pm0.1535$ & $52.71\pm11.62$ \\
EEG\mbox{-}TCNet~\cite{ingolfsson2020eegtcnet} & $44.52\pm11.86$& $0.1678\pm0.1779$ & $40.53\pm13.40$ & $46.25\pm10.71$& $0.1936\pm0.1415$  & $41.65\pm11.26$ & $52.66\pm9.12$ & $0.2899\pm0.1368$ & $48.15\pm11.36$ \\
EEGConformer~\cite{song2022eegconformer}       & $48.55\pm8.62$ & $0.2282\pm0.1295$ & $46.07\pm9.62$  & $50.23\pm9.61$ & $0.2534\pm0.1441$  & $48.01\pm10.81$ & $55.83\pm11.64$& $0.3374\pm0.1746$ & $53.82\pm12.81$ \\
BaseNet~\cite{wimpff2023basenet}               & $46.07\pm11.16$& $0.1911\pm0.1674$ & $41.80\pm13.52$ & $52.05\pm10.21$& $0.2960\pm0.1535$  & $48.59\pm12.55$ & $53.97\pm10.22$& $0.3095\pm0.1534$ & $50.14\pm12.05$ \\
LMDANet~\cite{miao2023lmdanet}                 & $46.13\pm10.70$& $0.1920\pm0.1604$ & $43.58\pm11.58$ & $47.89\pm8.87$ & $0.2184\pm0.1333$  & $45.78\pm10.46$ & $51.33\pm11.26$& $0.2701\pm0.1689$ & $49.03\pm12.17$ \\
STGENet~\cite{wang2024stgf}                    & $47.08\pm10.58$& $0.2034\pm0.1602$ & $43.01\pm12.60$ & $47.66\pm8.70$ & $0.2155\pm0.1253$  & $44.96\pm9.63$  & $50.83\pm10.28$& $0.2604\pm0.1552$ & $48.75\pm11.21$ \\
\textbf{STGMFM}                                & $\textbf{49.25}\pm\textbf{4.16}$ & $\textbf{0.2368}\pm\textbf{0.0643}$  & $\textbf{47.50}\pm\textbf{5.16}$  & $\textbf{57.26}\pm\textbf{8.34}$ & $\textbf{0.3592}\pm\textbf{0.1247}$ & $\textbf{56.52}\pm\textbf{8.08}$  & $\textbf{59.81}\pm\textbf{6.71}$ & $\textbf{0.3972}\pm\textbf{0.1006}$  & $\textbf{59.22}\pm\textbf{6.72}$ \\
\bottomrule
\end{tabular}
} 
\end{table*}

\subsection{Decision-level Fusion}
\label{subsec:fusion}
Let $[\cdot;\cdot]$ denote concatenation. We fuse logits with a shallow linear head,
\begin{equation}
  \mathbf{z}=\mathbf{W}\,[\mathbf{z}_A;\mathbf{z}_B;\mathbf{z}_C]+\mathbf{b},\qquad
  \hat{y}=\arg\max_k \mathbf{z}_A.
  \label{eq:fuse}
\end{equation}
We avoid gating/temperature mechanisms: in small EEG datasets they introduce extra degrees of freedom that are hard to calibrate across subjects and can overfit. A fixed linear combiner preserves class-wise alignment ability while remaining stable.

\subsection{Objective and Optimization}
\label{subsec:lossopt}
We train with cross-entropy and combine $\ell_1$ sparsity on graph-increment parameters with $\ell_2$ weight decay (AdamW). The total loss is
\begin{equation}
  \mathcal{L}=\mathcal{L}_{\mathrm{CE}}+\lambda_s\|\Delta\tilde{\mathbf{A}}\|_1+\lambda_t\|\Delta\tilde{\mathbf{S}}\|_1+\beta\!\sum_{\theta}\|\theta\|_2^2.
  \label{eq:loss}
\end{equation}
Cosine-annealed learning rates promote smooth convergence and better generalization. 

\section{Experiments and Results}

\subsection{Dataset and Protocols}
We collected a dry-electrode MI-EEG dataset with a 23-channel cap at 250~Hz. 19 subjects each completed two sessions on different days; within each session, three repeated runs were recorded for a three-class MI task. We evaluate three realistic protocols: 
(i) \textbf{Cross-Session}: train on one session and test on the other session from the same subject; 
(ii) \textbf{Cross-Subject}: leave one subject out for testing, train on the remaining subjects; 
(iii) \textbf{Cross-Subject + Single-Session Fine-tuning}: pre-train on other subjects and adapt using one session of the target subject.
Trials are segmented using windows of 125 samples (0.5~s) with a stride of 125, yielding nine time slices per trial.
We train 1000 epochs with AdamW (initial learning rate $2\!\times\!10^{-3}$), batch size 32, dropout 0.2, cosine-annealing schedule, and L1/L2 regularization on graph weights and classifier. 
All experiments run on a NVIDIA RTX~4090.

\subsection{Baselines and Overall Performance}
We compare against some representative models (
ShallowNet~\cite{schirrmeister2017deep}, 
EEGNet~\cite{lawhern2018eegnet}, 
EEG\mbox{-}TCNet~\cite{ingolfsson2020eegtcnet}, 
EEGConformer~\cite{song2022eegconformer},
BaseNet~\cite{wimpff2023basenet}, 
LMDANet~\cite{miao2023lmdanet}, 
STGENet~\cite{wang2024stgf}.)
We report 3 averaged metrics over 19 subjects: Accuracy, Cohen’s kappa and F1.
Table~\ref{tab:merged_main} summarize the results. 
STGMFM attains the best accuracy across all three protocols.

\begin{table}[t]
\centering
\caption{Ablation study on dry-EEG under Cross Subject. We report Avg$\pm$Std results for Accuracy (ACC), Cohen's kappa, and F1. Best results are \textbf{bold}.}
\label{tab:ablation}
\resizebox{\linewidth}{!}{%
\begin{tabular}{lccc}
\toprule
Variant & ACC(\%) & kappa & F1(\%) \\
\midrule
Only A\&B  & $54.19\pm8.87$ & $0.3132\pm0.1327$ & $53.53\pm8.78$ \\
Only C  & $51.14\pm9.16$ & $0.2669\pm0.1376$ & $50.11\pm9.45$ \\
No PLV initialization & $56.76\pm8.71$ & $0.3517\pm0.1278$ & $56.11\pm9.03$ \\
Spatial adjacency fixed & $52.20\pm9.48$ & $0.2834\pm0.1420$ & $51.64\pm9.50$ \\
No L1/L2 regularization & $54.32\pm9.46$ & $0.3153\pm0.1416$ & $53.56\pm9.38$ \\
Gated fusion & $52.89\pm9.43$ & $0.2936\pm0.1412$ & $52.22\pm9.35$ \\
\textbf{STGMFM (full)} & $\textbf{57.26}\pm\textbf{8.34}$ & $\textbf{0.3592}\pm\textbf{0.1247}$ & $\textbf{56.52}\pm\textbf{8.08}$ \\
\bottomrule
\end{tabular}
} 
\end{table}

\subsection{Ablation and Component Analysis (on Cross-Subject)}
We quantify the contribution of each component using the Cross-Subject protocol. 
Table~\ref{tab:ablation} reports the results.

\textbf{Modules Analysis:}
\emph{(i) Dual graph orders (A: CCG$\!\rightarrow\!$TSG; B: TSG$\!\rightarrow\!$CCG)}:
The two orders inject complementary inductive biases that hedge against distinct error pathways in dry-EEG: A first stabilizes spatial connectivity then consolidates temporal relations while B emphasizes slice-wise temporal consistency. 
\emph{(ii) Multi-Scale Frequency-Mixer (C)}:
MFM focuses on multi-scale envelopes and long-range modulations with reduced phase sensitivity, recovering ERD/ERS-like patterns that are blurred by dry-electrode noise.
\emph{(iii) PLV-initialized, learnable graphs}:
PLV provides physiologically plausible priors that narrow the search space and improve early-epoch stability, yet learnability preserves subject/session adaptability. At the same time, it also provides more detailed and precise adjacency.
\emph{(iv) Regularization and simple fusion}:
L1 prevents spurious edges/feature dominance, while L2 and cosine annealing smooth optimization.

\section{Conclusion}
We introduced STGMFM, a tri-branch fusion network for dry-electrode MI-EEG. Two complementary graph orders (CCG$\rightarrow$TSG and TSG$\rightarrow$CCG) and a Multi-Scale Frequency-Mixer jointly model spatio-temporal dependencies and envelope regularities. With PLV-initialized learnable graphs, cosine-annealed training, and L1/L2 regularization, STGMFM surpasses strong CNN/Transformer/GCN baselines across cross-session, cross-subject, and pre-train \& adapt protocols on a 23-channel dry-EEG dataset. Ablations confirm the role of each module under low-SNR, contact variability: dual graph orders curb noise propagation, MFM captures robust ERD/ERS-like envelopes, and simple decision-level fusion generalizes better than gated schemes. Future work will explore adaptive PLV estimation, subject-aware fusion, and channel pruning for on-device inference.

\vfill\pagebreak
\clearpage
\bibliographystyle{IEEEbib}
\bibliography{refs}

\end{document}